\pdfoutput=1

\documentclass[11pt]{article}
\usepackage{graphicx}
\usepackage{pdflscape}
\usepackage{ACL2023} 

\usepackage{times}
\usepackage{latexsym}
\usepackage{float}
\usepackage[T1]{fontenc}
\usepackage{multirow}
\usepackage[utf8]{inputenc}

\usepackage{microtype}

\usepackage{inconsolata}

\title{When Large Language Models Meet Citation: A Survey}
 
\author{{Yang Zhang$^{1,2}$\thanks{\emph{*Corresponding Author:yang.zhang@mq.edu.au}}}, {Yufei Wang}$^{1}$, {Kai Wang}$^{3}$, {Quan Z. Sheng}$^{1}$, {Lina Yao}$^{4,5}$\\   \textbf{{Adnan Mahmood}$^{1}$}, \textbf{{Wei Emma Zhang}$^{6}$}, \textbf{{Rongying Zhao}$^{2}$}
\\
            $^{1}$ School of Computing, Faculty of Science and Engineering, Macquarie University, Sydney, Australia.                 
           \\ 
              $^{2}$ School of Information Management, Wuhan University, Wuhan,  China.
              \\
              $^{3}$ School of Computer Science and Engineering, Nanyang Technological University, Singapore.
                           \\
              $^{4}$ CSIRO's Data61, Sydney, Australia.
               \\
              $^{5}$ Computer Science and Engineering, University of New South Wales, Sydney, Australia.
                           \\
             $^{6}$ School of Computer and Mathematical Sciences, The University of Adelaide, Adelaide, Australia.
             \\
}

\begin{document}
\maketitle
\begin{abstract}
Citations in scholarly work serve the essential purpose of acknowledging and crediting the original sources of knowledge that have been incorporated or referenced. Depending on their surrounding textual context, these citations are used for different motivations and purposes. 
Large Language Models (LLMs) could be helpful in capturing these fine-grained citation information via the corresponding textual context, thereby enabling a better understanding towards the literature. Furthermore, these citations also establish connections among scientific papers, providing high-quality inter-document relationships and human-constructed knowledge. 
Such information could be incorporated into LLMs pre-training and improve the text representation in LLMs. Therefore, in this paper, we offer a preliminary review of the mutually beneficial relationship between LLMs and citation analysis. Specifically, we review the application of LLMs for in-text citation analysis tasks, including citation classification, citation-based summarization, and citation recommendation. We then summarize the research pertinent to leveraging citation linkage knowledge to improve text representations of LLMs via citation prediction, network structure information, and inter-document relationship. We finally provide an overview of these contemporary methods and put forth potential promising avenues in combining LLMs and citation analysis for further investigation.

\end{abstract}

\section{Introduction}
Citations facilitate the exchange of specialized knowledge by enabling authors and readers to make targeted references across multiple contexts at the same time. A better understanding of the textual context associated with citations could lead to a precise and rich scientific literature network. With the emergence of Large Language Models (LLMs), such as ChatGPT/GPT4\footnote{\url{https://chat.openai.com/}} and PaLM2\footnote{\url{https://ai.google/discover/palm2}}, there is a growing interest in collaboratively leveraging the capabilities of LLMs and citations to mutually enhance one another.

\begin{figure}
    \includegraphics[width=\columnwidth]{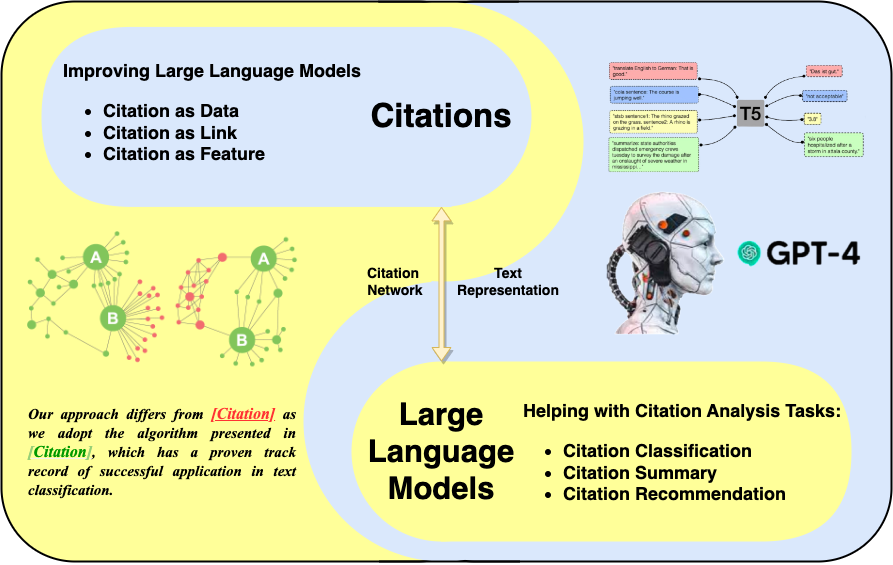}
    \caption{The mutually beneficial relationship between Large Language Models (LLMs) and citations.}
    \label{fig:LLM_citation}
\end{figure}

\begin{figure*}
    \includegraphics[width=1.0\textwidth]{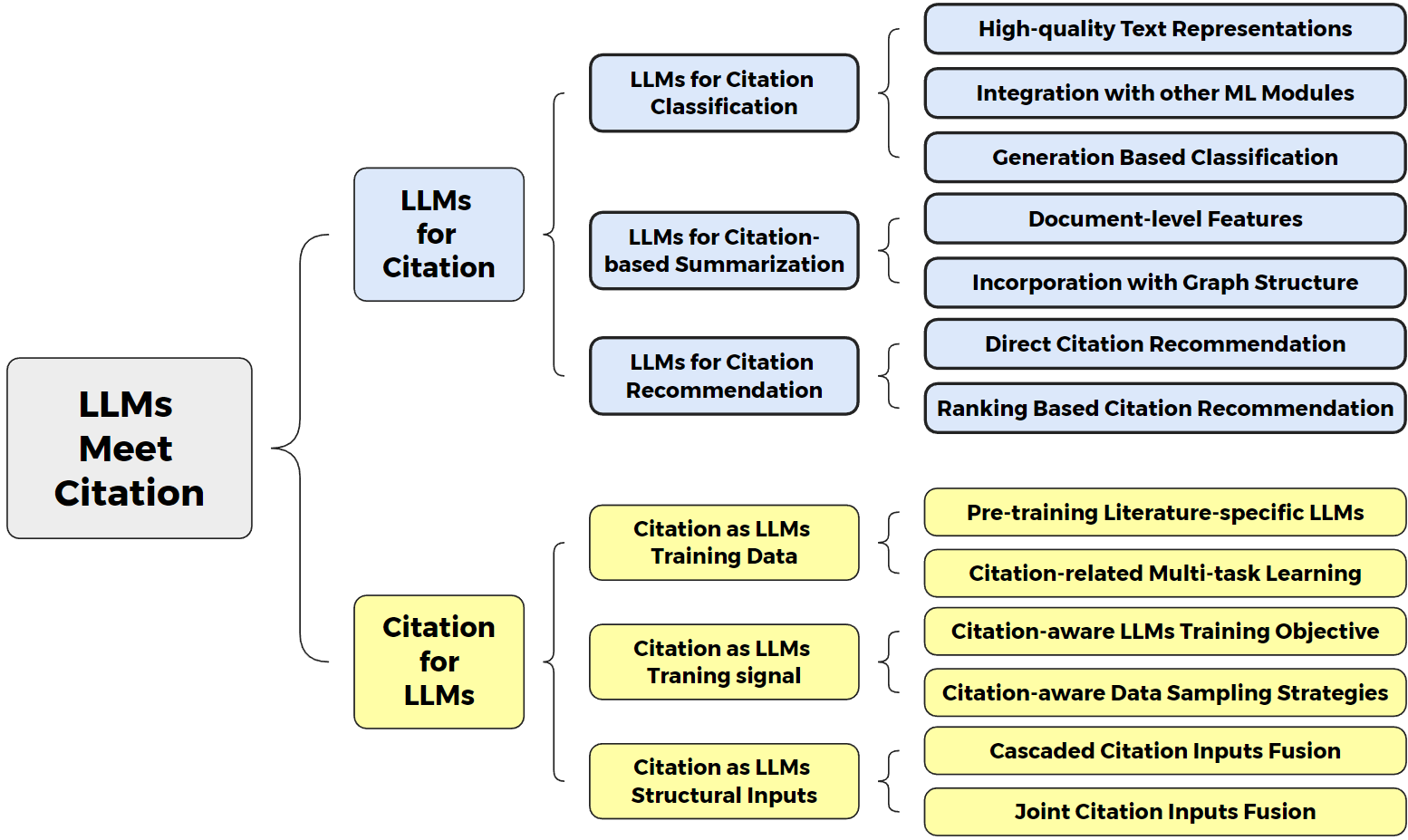}
    \caption{A taxonomy of the research for Large Language Models (LLMs) and citations.}
    \label{fig:LLM_citation_taxonomy}
\end{figure*}

As depicted in Figure~\ref{fig:LLM_citation}, there exists a mutually beneficial relationship between LLMs and citations. On the one hand, LLMs can provide high-quality textual features for understanding citation textual context. In this paper, we view LLMs as large Transformer-based language models trained on
massive document corpora using various self-supervised objectives. 
Depending on the architectural structure, LLMs can be categorized into a) \emph{Encoder-only LLMs} (e.g., BERT); b) \emph{Decoder-only LLMs} (e.g., GPT series) and c) \emph{Encoder-decoder LLMs} (e.g., T5 series). 
They can 
provide high-quality, dense vectors for text and generate coherent and fluent natural language. 
Empirically, LLM-based models could effectively comprehend citation textual context and produce high-quality citation analysis results. 
On the other hand, citations could establish high-quality literature linkage in LLM pre-training corpus. Such linkage revealing multi-hop knowledge and information spanning across multiple documents is largely overlooked by many existing LLMs. 
By integrating the continuously growing and up-to-date citation network, LLMs would learn how humans acquire knowledge and make discoveries across different but relevant sources~\citep{DBLP:journals/corr/abs-2304-10668}.

\begin{table*}[!ht]
\caption{Previous Surveys of Citation Related Tasks}
\footnotesize
\begin{tabular}{cccl}
\hline
Article Year & Research Method & Research Domain   & \multicolumn{1}{c}{Review Focus}                                                                                                                                                                                                                                 \\ \hline
\citeyear{ding2014content}       & Non-systematic  & Bibliometrics    & \begin{tabular}[c]{@{}l@{}}A survey on the foundations, methodologies, and applications of \\ citation content analysis, providing an overview of the current state \\ of the art in the citation function analysis.\end{tabular}                            \\
\citeyear{hernandez2016survey}       & Non-systematic  & Computer Science & \begin{tabular}[c]{@{}l@{}}This paper presents a summary of current research in Natural \\ Language Processing-driven citation analysis, highlighting the \\ experiments and practical examples that demonstrate the significance \\ of this field.\end{tabular} \\
\citeyear{jha2017nlp}       & Non-systematic  & Computer Science & \begin{tabular}[c]{@{}l@{}}The authors summarize studies that have addressed the issue of \\ identifying and classifying citation context. They examine the \\ most recent techniques and data repositories used for citation \\ context analysis.\end{tabular}   \\
\citeyear{lyu2021classification}       & Systematic      & Bibliometrics    & \begin{tabular}[c]{@{}l@{}}This study employed a meta-synthesis approach to develop a new \\ comprehensive classification of citation motivations, based on \\ previous research.\end{tabular}                                                                   \\
\citeyear{iqbal2021decade}       & Systematic      & Computer Science & \begin{tabular}[c]{@{}l@{}}The primary focus of this survey is on publications that employ \\ natural language processing and machine learning techniques \\ to analyze citations.\end{tabular}                                                                  \\ \hline
\end{tabular}
\label{survery_compare}
\end{table*}

Motivated by this remarkable co-development connection between LLMs and citations, how to effectively leverage the unique strengths from both sides has emerged as a promising research direction. There are a few studies that attempt to review relevant papers in this domain. However, they either only discuss the citation analysis tasks setup and designs~\cite{ding2014content,lyu2021classification} or simply focus on the general machine learning and NLP technologies applications on citation analysis tasks~\cite{jha2017nlp,iqbal2021decade}. There are limited research efforts summarizing how state-of-the-art LLMs help with citation analysis tasks. Furthermore, none of these studies particularly look into the mutually beneficial relationship between LLMs and citations (i.e., how LLMs could further benefit from citation-relation data). We refer readers to Table~\ref{survery_compare} for more details on previous citation-related surveys. 

To address this disparity, we undertake an initial survey to explore how LLMs and citation analysis tasks can derive mutual benefits from each other. Figure~\ref{fig:LLM_citation_taxonomy} shows the taxonomy of this survey. We first review how LLMs contribute to the in-text citation analysis tasks, including citation classification, citation summarization, and citation recommendation. In general, the strong text representations from LLMs are helpful in modelling the complex textual context around the citations. We then examine how LLMs could benefit from various citation corpus. This includes directly using citations as LLMs training data, incorporating citations as LLMs training signals, and leveraging citations as LLMs structural inputs. Incorporating information from multiple documents helps LLMs to learn multi-hop knowledge that does not exist in single-document training data\cite{DBLP:journals/corr/abs-2305-11070}. Finally, we provide an overview of future research opportunities in tightening the connections between citations and LLMs.

\section{LLMs for Citation}
\label{llm4citation}

\begin{figure}
\includegraphics[width=\columnwidth]{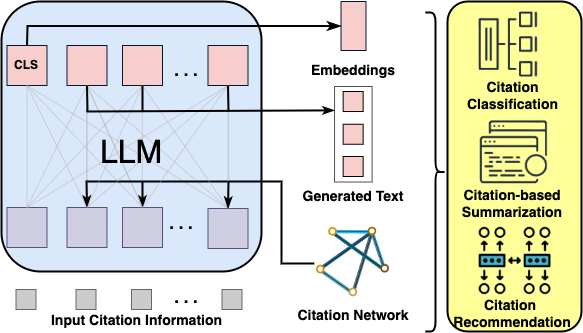}
    \caption{Large Language Models (LLMs) improve citation-related tasks via their high-quality text embedding, strong text generation capability and feasibility to incorporate citation structure information.}
    \label{fig:LLM_for_citation}
\end{figure}

In this section, we discuss how 
LLMs improve the performance of various citation-related tasks, including citation classification, citation-based summarization, and citation recommendation. As shown in Figure~\ref{fig:LLM_for_citation}, LLMs achieve this goal by providing high-quality text embedding, strong text generation capability, and flexibility to incorporate citation structure information.

\subsection{LLMs for Citation Classification} 
\label{llmcitationcls} 
Consistent with earlier studies, we perceive citation classification task as comprising two unique components: classification based on citation intent and citation sentiment.\cite{teufel2006automatic, abu2013purpose, zhang2021tdm}.


Following~\cite{teufel2006automatic, zhang2021tdm}, we also consider Citation sentiment as part of the citation function in the citation classification task. The task of citation classification involves assigning a functional label (e.g., ``Background Information'', ``Method'' or ``Positive'') to a specific citation instance. This is done by providing inputs (e.g., textual citation context and citing sentences) to the classification model. The fundamental idea is to determine the citing reasons and sentiment via their surrounding textual context. 


Given the time and cost constraints associated with manual citation function labeling, and the growth in scientific literature, the need for automated citation function classification, typically seen as a single-label classification task, has surged ~\cite{iqbal2021decade}.
Various sophisticated deep learning models have been applied to this task~\cite{cohan2019structural}. 
In recent times, the employment of LLMs has gained significant traction in the realm of citation function classification tasks~\cite{roman2021citation}. This is largely attributed to the superior performance these models deliver, courtesy of their impressive prowess in handling text classification tasks. 
\paragraph{High-quality Text Representations.}
Incorporating LLMs, e.g., BERT~\cite{devlin2019bert}, SciBERT~\cite{beltagy2019scibert}, and T5~\cite{raffel2020exploring}, into citation function classification tasks has shown considerable advantages. These models excel at achieving superior text representation, thereby significantly aiding the enhancement of citation classification task performance~\citep{maheshwari2021scibert, visser2022sentiment, lauscher2021multicite, roman2021citation, te2022citation}. For example,~\citet{visser2022sentiment} assesses the use of BERT for citation classification of scientific literature by leveraging the ``[CLS]'' as the output vector and further fine-tuning the citation classification model to obtain better performance.

\paragraph{Integration with other ML Modules.}
LLMs can not only provide high-quality text representation along but also integrate other machine learning modules. For instance,~\citet{budi2023understanding} propose a multi-output convolutional neural network model based classifier, and they use BERT as the initial word embedding encoder to encode the input text and obtain better performance of citation classification task.~\citet{wang2020gbdt} propose a hybrid approach that combines the Gradient Boosting Decision Tree (GBDT) model with the BERT model. They utilize linear blending to harmonize the stacking models and the BERT model, thereby producing their ultimate outcomes.

\paragraph{Generation Based Classification.}
Above LLMs are commonly used as \emph{discriminative LLMs} (i.e., encoder-only) that can 
produce hidden representations to the input text. There are also \emph{Generative LLMs} (i.e., Seq2Seq or Decoder-only) that are capable of directly generating high-quality text. Recently,~\citet{zhang2022towards} propose a novel generative citation function classification model based on T5~\cite{raffel2020exploring}. Instead of mapping neural hidden states to label probability, they directly train the model to generate the citation function labels (e.g., Motivation) at the decoder side for the given inputs. Such generative LLMs can also be high-quality synthetic data generators.     ~\citet{gupta2023inline} combine the GPT-2 into their framework to obtain the high-quality synthetic data for citation function classification task. In the same trend,~\citet{zhang2023hybrid} propose HyBridDA, which combines the retrieved results from the large unlabelled corpus and fine-tunes the GPT-2 model to generate desired and novel citation sentences to tackle the data sparsity and data imbalance issue. The generated syntactic data are then combined with existing real training data to fine-tune SciBERT as their final classification model. 
\subsection{LLMs for Citation‑based summarization}
Aggregating all citation sentences referring to a paper can yield a significant amount of information about the paper itself~\cite{abu2011coherent}.
Citation-based summarization harnesses the citations from a scientific article or its reference citation network to assemble a succinct summary. The summary aims to highlight the main contributions of papers, based on
citation sentences from papers that cite the source document. The initial attempt uses sentence clustering and ranking methods~\cite{luo2023citationsum}.

With the progression of research, automatic citation-based summarization can be broadly divided into two distinct categories as follows: (a) document-level citation context span based summarization~\cite{qazvinian2008scientific}. Techniques involving span identification and varied feature extraction have been utilized~\cite{cohan2015scientific};
(b) citation network structured aware summarization. More recent research works put forward the idea that citation graphs can be instrumental in creating top-notch summaries of scientific papers~\cite{chen2022comparative, luo2023citationsum}. 
LLMs, with their proficiency in deriving complex text representations from dense vectors and generating high-quality text, are gaining prominence in complex tasks like citation-based summarization. Hence, there's an enhanced capability to encode citation context and integrate citation network graphs, thereby further boosting citation-based summarization efforts~\cite{luo2023citationsum}. 


\paragraph{Document-level Features.}

Given their superior capability to derive text representation from textual input, LLMs can be employed for capturing document-level features within a citation recommendation system.~\citet{saini2023multi} introduce a multi-view clustering approach that utilizes dual-embedding spaces and citation contexts for document summarization, leveraging document encoders like BERT, GPT-2, and XLNET to capture document-level features.~\citet{roy2022biocite} propose a method for establishing connections between citing sentences and corresponding citation sentences in the referred literature.
Their framework pairs every sentence in a publication with all sentences from the referenced document, then seeks to identify semantically similar pairs. The selected sentences from the referenced paper, which have semantic similarities with the publication's sentences, are considered cited statements. This process entails the development of a citation linkage framework using both sequential and tree-structured models based on Bio-RoBERTa. 
In addition,~\citet{mao2022citesum} feed document-level text, i.e., citation text, abstract, and conclusion into the generative models BART-large~\cite{lewis2020bart} and PEGASUS-large~\cite{zhang2020pegasus} as training data, using the subsequent trained models to effectively produce comprehensive scientific extreme summarization, i.e. single sentence summaries.

\paragraph{Incorporation with Graph Structure.}
The above citation-based summarization methods only consider surface document information. However, relevant citations can provide rich literature network information. Thus, some research propose to combine LLMs with Graph-based neural networks to effectively incorporate these additional information.~\citet{chen2022comparative} first shift through comparative citations to identify objects of comparison. They then construct a comparative scientific summarization corpus and put forward the Comparative Graph-based Summarization BERT (CGSUM-BERT). This model is designed to craft comparative summaries with the assistance of citation guidance.~\citet{cai2022covidsum} propose a model called COVIDSum for summarizing COVID-19 scientific papers, which leverages a linguistically enriched SciBERT framework. The process begins with the extraction of key sentences from the source papers and the construction of word co-occurrence graphs. A SciBERT-based sequence encoder is then employed to encode the sentences, while a Graph Attention Networks-based graph encoder processes the word co-occurrence graphs. In the final stage, these two encodings are fused to generate an abstract summary of each scientific paper.

\subsection{LLMs for Citation Sentence Generation}
The emerging and valuable task of generating citation sentences independently is made possible through the capabilities of LLMs.
~\citet{xing2020automatic} initially train an implicit citation text extraction model using BERT. They utilize this model to create an extensive training dataset for the citation text generation task. Subsequently, they introduce and train a multi-source pointer-generator network equipped with a cross-attention mechanism for citation text generation.
BACO~\cite{ge2021baco} employs its encoder to encode both the citation network and the text from both citing and cited papers. BACO's generator assesses the significance of sentences within the cited paper's abstract and leverages this knowledge for text generation. Moreover, the framework undergoes additional training with citation functions.
Both of them are dedicated to the generation of citation sentences for multiple references while also incorporating intent control.

\subsection{LLMs for Citation Recommendation}
Citation recommendation serves as an invaluable tool for academic publishers and scholars~\cite{pillai2022survey} by producing suitable references that align with the specified query text~\cite{iqbal2021decade}. ~\citet{taylor2022galactica} constitutes the most extensive and comprehensive endeavor to pre-train literature-specific LLMs. Additionally, the Galactica paper introduces a novel citation token for pre-training and explores its influence on the citation recommendation task.

In a formal context, a citation recommendation model is like a helpful tool that suggests a selection of academic papers that researchers might find useful.
Suppose we have a collection of papers and a group of researchers. The citation recommendation model
evaluates how beneficial a specific paper from this collection could be to a particular researcher. This evaluation is done by a so-called utility function that measures the benefit. In other words, it determines how well a paper matches a researcher's needs or interests. The model then recommends a paper that maximizes this benefit for each researcher. This is typically reflected in the form of a rating given by the researchers to the suggested papers.

Conventionally, various components of citation recommendation systems heavily relied on traditional machine learning models~\cite{yasunaga2019scisummnet}. However, the recent surge in popularity of Large Language Models (LLMs) in these systems is primarily due to their exceptional text comprehension capabilities. 
We observe that the use of LLMs in these aforementioned types of citation recommendations bears similarities.
\paragraph{Direct Citation Recommendation.}
Direct citation recommendation involves picking the best-suited item from a collection of potential recommendations. By converting unstructured text data into high-dimensional, meaningful vectors, LLMs bolster citation recommendation systems' ability to thoroughly analyze and utilize the inherent substance of scientific papers, thereby enhancing recommendation efficacy. For instance,~\citet{jeong2020context} introduces a deep learning model and a well-structured dataset designed for context-aware paper citation recommendation. Their model, which consists of a document encoder and a context encoder, utilizes both Graph Convolutional Networks layer and BERT to directly predict the label of the recommended citation. The model is trained using cross-entropy as the loss function.
In a similar way,~\citet{bhowmick2021augmenting} infuse their citation recommendation system with citation context encoded by SciBERT, along with citation history and co-authorship data processed through GCN. This system aims to directly predict the probability of the label tied to the proposed output.~\citet{dai2023heterogeneous} also propose a novel neural network-based approach called Citation Relational BERT with Heterogeneous Deep Graph Convolutional Network (CRB-HDGCN) for the task of inline citation recommendation related to COVID-19. The model leverages a retrained BERT on an augmented citation sentence corpus and HDGCN to extract reliable vectors and subsequently predict potential citations directly.

\paragraph{Ranking Based Citation Recommendation.}
Ranking-based citation recommendation is to rank the corresponding citations based on the given textual inputs. LLMs' capacity to understand and generate high-quality text can add a level of abstraction that traditional machine learning models often lack, making these models more adaptable to varied and complex tasks. For example,~\citet{nogueira2020navigation} interprets citation recommendation as a ranking problem and propose to use a two-tier method involving initial candidate generation, followed by a re-ranking process. Within this structure, an effective combination tailored to the scientific domain is utilized, which includes a ``bag of words'' retrieval process, succeeded by a re-scoring phase using a BERT model. A more comprehensive method, Deepcite that uses a hybrid neural network and focuses on content is introduced by ~\citet{wang2020content}. Initially, the BERT model is used to pull out complex semantic meanings from the text. Then, to gain both the local and sequence context in sentences, they use a multi-scale CNN model and a Bi-LSTM model. This results in matching text vectors that create sets of potential citations. Finally, a deep neural network rearranges these sets in order of relevance, considering both their individual scores and other various features. Similarly,~\citet{gu2022local} investigate prefetching by utilizing nearest neighbor search within text embeddings, developed through a hierarchical attention network. They introduce a method named the Hierarchical Attention encoder (HAtten). When paired with a SciBERT reranker, fine-tuned on local citation recommendation tasks, it delivers high prefetch recall for a defined pool of candidates awaiting reranking. Also, ~\citet{abbas2022deep} proposed a simply but more effective method by employing Word2Vec and BERT models to represent their dataset through word embeddings, and introduce using Bi-LSTM based ranker to classify the rhetorical zones of research articles. Their system computes similarity using rhetorical zone embeddings to effectively tackle the cold-start problem. It is a prevalent occurrence for a single context to be substantiated by multiple co-citation pairs. To tackle this issue,~\citet{zhang2022mp} propose \emph{Multi-Positive BERT Model} (MP-BERT4REC) which is designed to align with a sequence of Multi-Positive Triplet objectives, thereby facilitating the recommendation of several positive citations in response to a specific query according to the similarity ranking. 
~\cite{kieu2021learning} put forward a unique approach to citation recommendation that capitalizes on Sentence-BERT's deep sequential document representation. This is combined with Siamese and triplet citation networks within a submodular scoring function. This scoring function aids in the selection of the most suitable candidates, taking into account a balance between their relevance. This method marks the first instance of combining LLMs and submodular selection for citation recommendation tasks.


\begin{figure}
\includegraphics[width=\columnwidth]{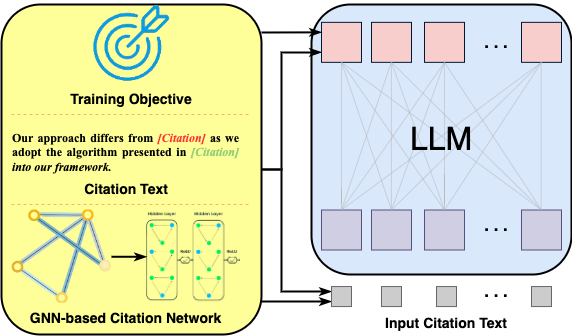}
    \caption{Citations are helpful and unique resources in improving and shaping LLMs. They can act as LLMs training objects, as LLMs training data and as LLMs structural inputs.}
    \label{fig:citation_for_LLM}
\end{figure}

\section{Citation for LLMs}
\label{citation4llm}
In previous sections, we discuss how LLMs help with various citation analysis tasks and provide rich and precise understandings about the deep semantics of citations. Interestingly, citations also establish high-quality literature linkage, potentially improving LLMs in understanding highly organized texts from multiple documents, and enabling their acquisition of multi-hop knowledge by covering diverse academic fields' research. As shown in Figure~\ref{fig:citation_for_LLM}, in this section, we summarize the research progress of leveraging citation linkage data to improve LLMs through different dimensions. According to the usage of citations, we divide recent relevant research into three categories: Citation as LLMs Training Data, Citation as LLMs Training Signals, and Citation as LLMs Structural Inputs.


\subsection{Citation as LLMs Training Data}

The most straightforward way to leverage citations for LLMs is to construct citations as training data for LLMs pre-training or fine-tuning. Some researchers focus on applying citation corpus to train LLMs, while others explore training LLMs via citation-related tasks.

\paragraph{Pre-training Literature-specific LLMs.}
The success of general LLMs, such as BERT, motivates further exploration of literature-specific LLMs.
As a core role in scientific literature data, the citation-related text is bound to participate in the pre-training process. 
Following the BERT language model, SciBert \cite{DBLP:conf/emnlp/BeltagyLC19} is pre-trained on a vast multi-domain corpus consisting of 1.14M scientific papers.
SsciBert \cite{DBLP:journals/scientometrics/ShenLLHZLFW23} is pre-trained on papers from Social Science Citation Index (SSCI) journals, compensating for the lack of LLMs on the asocial science domain. 
~\citet{DBLP:journals/health/GuTCLULNGP22} focus on the bio-medicine literature and propose PubMedBERT with the strategy of domain-specific pre-training from scratch on a PubMed corpus. 

\paragraph{Citation-related Multi-task Learning.}
In the multi-task pre-training, the LLM model is trained to perform all tasks with data annotated for different tasks.
As citation data creates rich structure information,
citation-related tasks also play a role in pertaining LLMs. One respective citation task, citation prediction, is to find a pair of citing and cited scientific papers.
To achieve multi-task pre-training involving citation prediction, some recent studies utilize the contrastive Learning technique to design pre-training objectives. 
Given a query paper $q$ and the positive and negative candidates, $c_+$ and $c_-$, a common contrastive learning objective is to pull $q$ and relevant $c_+$ together while pushing $q$ and irrelevant $c_-$ apart. Specifically, SciMult~\cite{DBLP:journals/corr/abs-2305-14232} is a novel multi-task LLM pre-training framework for scientific literature understanding tasks.
Treating citation prediction as a sub-task, SciMult focuses on facilitating common knowledge sharing across different tasks while preventing task-specific skill interference. To achieve it, the hard negative mining technique used in the citation prediction task is employed for scientific paper contrastive learning in different tasks~\cite{DBLP:conf/acl/CohanFBDW20}. 
Utilizing a batch-wise contrastive learning objective, ConGraT~\cite{DBLP:journals/corr/abs-2305-14321} is a self-supervised framework for simultaneously learning individual representations for texts and nodes within a parent graph, such as citation graphs in Pubmed with each paper related to a specific node. 
\citet{DBLP:journals/corr/abs-2212-03533} propose E5, a new family of text embeddings trained using weak supervision signals in a contrastive manner on a large text pair dataset, CCPairs. As an important data part of CCPairs, each instance from scientific literature includes a title, abstract, and a pair of citing and cited papers.

\subsection{Citation as LLMs Training Signals}
\label{citationlink}

Citations create rich and high-quality semantic links between scientific documents. However, the mainstream pre-training objectives of LLMs, such as Masked Language Modelling (MLM), purely encode texts and do not take inter-document structure information into consideration. To fill this gap, recent studies draw attention to document-level LLMs and propose new sampling strategies and pre-training objectives considering these document connections, especially citations.

\paragraph{Citation-aware LLMs Training Objective.}
Previous LLMs like BERT are aimed at token- and sentence-level training objectives. Some recent language models are proposed specifically for document-level representation, making it more suited for document-level downstream applications such as literature classification and citation recommendation. 
SPECTER~\cite{DBLP:conf/acl/CohanFBDW20} achieves document-level representation using citation-aware Transformers. 
It captures inter-document relatedness in the citation graphs, utilizes citations as an inter-document incidental supervision signal, and transforms this signal into a triplet-loss pre-training objective. 
\citet{DBLP:conf/sigir/RamanSV22} propose a novel document retrieval method that combines intra-document content with inter-document relations in learning document representations. Benefiting from the inter-document citation relationships, a contrastive learning-based quintuplet loss function is designed, which pushes semantically similar documents closer and structurally unrelated documents further apart in the representation space. Additionally, the model varies the separation margins between the documents based on the strength of their relationships.
PATTON~\cite{DBLP:journals/corr/abs-2305-12268} is specifically designed to handle text-rich networks which are made up of text documents and their interconnections, such as citations and co-authorships in an academic context. Unlike current pre-training approaches that primarily focus on texts without considering inter-document structural data, Patton integrates two pre-training strategies: network-contextualized masked language modeling and masked node prediction, to effectively capture the correlation between text attributes and network structures.

\paragraph{Citation-aware Data Sampling Strategies.}
A bias in pre-training language models has been revealed that the model creates stronger dependencies between text segments appearing in the same training example than those in separate examples \cite{DBLP:conf/iclr/LevineWJNHS22}. Thereby, some studies propose new text sampling and batching strategies considering document connections.
LinkBERT~\cite{DBLP:conf/acl/YasunagaLL22} treats a text corpus as a graph of interlinked documents and creates language model inputs that put linked documents within the same context. Then, the Bert-based language model is pre-trained with two self-supervised objectives, namely masked language modeling and the newly proposed document relation prediction.
SciNCL~\cite{DBLP:conf/emnlp/OstendorffRAGR22} is an LM-based approach for scientific document representation learning, which utilizes controlled nearest neighbor sampling over citation graph embeddings for contrastive learning. The method introduces the concept of continuous similarity, enabling the model to recognize similarities between scientific papers even without direct citation links, and avoids collisions between negative and positive samples by controlling the sampling margin between them.


\subsection{Citation as LLMs Structural Inputs}

The above approaches simply use pure textual data as the inputs to LLMs and treat citation graphs as data sources or training objectives. However, they could potentially overlook the complex and fine-grained citation graph structure information. With the rapid development of Graph Neural Networks (GNNs), recent researchers propose to integrate GNNs into LLMs by fusing the node representation of citation graphs, facilitating better understandings of scientific literature as well as natural language. We categorize previous research into \emph{Cascaded Citation Inputs Fusion} and \emph{Joint Citation Inputs Fusion}. 


\paragraph{Cascaded Citation Inputs Fusion.}
Some researchers propose to infuse citation graph inputs using a cascaded model architecture where LLMs independently encode the textual input as embeddings and then GNNs are applied to amalgamate these embeddings.~\citet{Guan2022AHM} presents Citation Graph Collaborative Filtering (CGCF), which is a combination of document representation and Graph Neural Network, to improve automated recommendation systems for scientific articles. 
The method begins with building a user-paper bipartite graph based on citation relations, then initializes the paper's embedding using its title and abstract through a pre-trained language model. The final step refines node embeddings by GNNs to achieve a more holistic and in-depth representation of user-paper interactions and semantic content.
\citet{DBLP:conf/aaai/WangSLCJ0W22} propose DisenCite, a novel disentangled representation-based model for automatically generating citation text by integrating paper text representation from LMs and citation graph representation from GNNs. Unlike prior approaches, this method produces context-specific citation text, allowing for the generation of different citation types for the same paper. 

\paragraph{Joint Citation Inputs Fusion.}
The above architecture is marked by a key limitation: it models textual features in isolation, which detracts from its effectiveness. Therefore, some scholars further explore various novel architectures to combine citation graphs and LLMs.
GraphFormers~\cite{DBLP:conf/nips/YangLXLLASSX21} integrates Graph Neural Networks (GNN) components into the transformer blocks of language models for more efficient textual graph representation learning. i.e. they demonstrated the model was effective on the DBLP citation prediction task. Unlike existing works, GraphFormers blends text encoding and graph aggregation into an iterative workflow, thereby comprehending each node's semantics from a global perspective. Furthermore, the authors propose a progressive learning strategy where the model is incrementally trained on manipulated and original data to enhance its ability to integrate information on a graph.
\citet{DBLP:journals/corr/abs-2304-10668} present a novel Graph-Aware Distillation framework (GRAD) to address the scalability issue in combining Graph Neural Networks (GNNs) with Language Models (LMs) by using citations graphs. Unlike traditional knowledge distillation, GRAD concurrently optimizes a GNN teacher model and a graph-free student model via a shared LM, which allows the student model to use the graph information encoded by the teacher model. 
\citet{DBLP:journals/corr/abs-2305-11070} enhance deep learning language models by incorporating graph-represented information from citation graphs with an early fusion strategy, where the GNN component has access only to the graph and to the static textual data, while BERT is fed directly with both textual and graph information.

\section{Future Directions}\label{futher}
In the previous sections, we have reviewed how LLMs and citations mutually benefit from each other. However, there are still many challenges and open problems that need to be addressed. In this section, we discuss the future directions in further tightening LLMs and citations.

\subsection{Citation for Responsible and Accountable LLMs}
Large Language Models (LLMs) offer revolutionary advantages but also present distinctive challenges, particularly in the domains of intellectual property (IP) and ethical considerations. 
~\citet{DBLP:conf/nips/BrownMRSKDNSSAA20} takes a novel approach to address these challenges by drawing parallels between LLMs and well-established web systems by using citation as the essitail component in LLMs. Furthermore, they suggest that a comprehensive citation system for Large Language Models (LLMs) should encompass both non-parametric and parametric content. To steer future endeavors aimed at constructing more responsible and transparent Large Language Models (LLMs), a set of research challenges in this domain should be considered.

\subsection{Zero-shot/Few-shot Learning for Citation-related Tasks}
As discussed in Sec.~\ref{llm4citation}, many LLM-based approaches heavily rely on a large amount of labeled training data, ranging from a few hundred to a few thousand. 
However, recent state-of-the-art LLMs, such as GPT-3~\cite{DBLP:conf/nips/BrownMRSKDNSSAA20} and ChatGPT, have demonstrated superior zero-shot and few-shot ability in handling various NLP tasks. 
It is interesting to explore the strengths and weaknesses of the existing LLMs when handling citation-related tasks e.g, citation function classification and citation sentiment classification.  
As general LLMs are normally trained with very limited academic training corpora, one could verify whether the LLMs trained with a sufficiently large portion of academic corpora are equipped with stronger zero/few-shot ability than these general LLMs.  

\subsection{Citation Instruction Tuning}
Recently, instruction tuning that distils various skills and knowledge from existing strong LLMs~\cite{wei2022finetuned,sanh2022multitask,DBLP:conf/nips/Ouyang0JAWMZASR22,wang-etal-2022-super,DBLP:journals/corr/abs-2211-13308}, has received significant research interests.
Recently,~\cite{DBLP:journals/corr/abs-2305-14232} propose to unify three citation-related tasks as a set of unified instructions for better learning multi-task knowledge. It will be interesting to further unify a diverse set of citation-related tasks as instructions and explore whether fine-tuning the LLMs with these instructions can further improve the performance of these LLMs and generalize well on those unseen tasks.


\subsection{Expanding Citation Networks For LLMs}
In Sec.~\ref{citation4llm}, we only consider incorporating citation text into LLMs. However, the citation is not the only type of data that builds linkage between different text (e.g., papers, documents, web pages)~\cite{DBLP:journals/sigkdd/GetoorD05a}. It will be interesting to expand the existing citation network to other types of data sources. For example, linking scientific papers to relevant knowledge graph entities or social media posts that talk about similar topics or papers. Such an expanded graph includes more diverse textual information and should be helpful in training or fine-tuning high-quality LLMs.

\subsection{LLMs-based In-text Citation Analysis}
While the primary focus of this paper revolves around the realms of Natural Language Processing (NLP) and Machine Learning (ML), it also encompasses discussions that draw from bibliometrics journals, such as Scientometrics, thereby fostering an interdisciplinary dialogue. For instance, consider a bibliometric study within the context of Large Language Models (LLMs). An illustrative example could involve the characterization of citation distribution emanating from GPT-4 or the examination of potential biases in citations within generative LLMs. Notably, tasks like citation sentiment classification serve as essential precursor steps in the analysis undertaken by bibliometric researchers.

\section{Conclusion}
\label{conclusion}
This paper pioneers a summary of existing interdisciplinary research on how Large Language Models (LLMs) contribute to citation-related tasks. We engage in a comprehensive study exploring how citation network and citation semantic information enhance LLMs for superior text representation. We then encapsulate the shortcomings of current research and put forth several potential research pathways. Our efforts could potentially be beneficial for researchers engaged with LLMs and citation-related tasks.

\section{Limitation}

A constraint of our study is its exclusive focus on primary studies published in English, neglecting significant research on citation functions published in other languages such as Chinese.


\bibliography{anthology, custom, citation4llm}
\bibliographystyle{acl_natbib}

\end{document}